\documentclass[referee,useAMS,usenatbib]{mn2e}
\usepackage{graphicx}
\usepackage{hyperref}
\usepackage{epstopdf}
\usepackage{lscape}
\usepackage{bm}

\title[Mass and orbital parameters : HD~213597B]
{Determination of mass and orbital parameters of a low-mass star
HD~213597B}

\author[Chaturvedi et al.]
  {Priyanka Chaturvedi$^{1}$\thanks{E-mail:priyanka@prl.res.in},
   Rohit Deshpande$^{2,3}$,
   Vaibhav Dixit$^{1}$,
   Arpita Roy$^{2,3}$
\newauthor
   Abhijit Chakraborty$^{1}$,
   Suvrath Mahadevan$^{2,3}$,
   B.G. Anandarao$^{1}$,
\newauthor
   Leslie Hebb$^{4}$ and
   P. Janardhan$^{1}$
\\
$^1$ Astronomy \& Astrophysics Division, Physical Research Laboratory, Ahmedabad 380009, India\\ 
$^2$ Dept. of Astronomy \& Astrophysics, Pennsylvania State University, University Park, PA 16802\\
$^3$ Center for Exoplanets and Habitable Worlds, The Pennsylvania State University, University Park, PA 16802\\
$^4$ Department of Physics, Hobart and William Smith Colleges, Geneva New York 14456, USA\\}

\date{Received 2014 XXXX XXX }

\pagerange{\pageref{firstpage}--\pageref{lastpage}} \pubyear{2014}

\def\LaTeX{L\kern-.36em\raise.3ex\hbox{a}\kern-.15em
    T\kern-.1667em\lower.7ex\hbox{E}\kern-.125emX}

\begin{document}
\label{firstpage}

\maketitle

\begin{abstract}

HD~213597 is an eclipsing binary system which was detected by the \textit{STEREO} spacecraft and was speculated to host a low-mass stellar companion. We used high-resolution spectroscopy with the 10-m Hobby-Eberly Telescope and the 1.2-m telescope in Mt Abu for radial velocity (RV) measurements of this source. We performed aperture photometry for this star on the \textit{STEREO} archival data and thereby confirm the transit signature. We also did follow-up ground-based photometry with a 10-inch telescope from Mt Abu. The spectroscopic RV semi-amplitude of the primary (33.39~km~s$^{-1}$) indicates that the secondary is an M dwarf making the system a short period F+M eclipsing binary. These RVs along with the inclination derived from our combined photometric analysis ($i~=~84\degr.9$), enable us to estimate the mass of the secondary as $M_{B}~\sim~0.286~M_{\sun}$ and radius as $R_{B}~\sim~0.344~R_{\sun}$ using an estimated mass ${M_{A}~\sim~1.3~M_{\sun}}$ and radius ${R_{B}~\sim~1.97~R_{\sun}}$ of the primary. Our spectral analysis returned the following parameters: $T_{\rm{eff}}$~=~6625$~\pm~121$~K, $[Fe/H]$~=~$-0.095~\pm~0.08$ and ${\rm log}~g$~=~$3.72~\pm~0.22$ for the primary. When ${\rm log}~g$ is constrained to a value of $3.96$, we derive $T_{\rm{eff}}$~=~6753$~\pm~$52K and $[Fe/H]$~=~$-0.025~\pm~0.05$.
\end{abstract}


\section{Introduction}
Mass and radius are the two fundamental properties of a star that allow us to determine their age, evolution and luminosity \citep{andersen91}. Through careful photometry and spectroscopy of detached single and double-lined eclipsing binaries, it is possible to obtain radii and masses of individual components to as high accuracies as 1 per cent for double-lined eclipsing binaries. Such precise measurements impose strong constraints on stellar evolution models. In recent years, such measurements have revealed discrepancies between observed and model-derived stellar radii of low-mass stars \citep{Torres2010}. For stars with masses below $0.5~M_{\sun}$, the measured radii appear to be 20--30 per cent larger than predicted \citep{Lopez-Morales2007}. One hypothesis suggests that this disagreement is caused by the degree of magnetic activity in stars: strong magnetic fields inhibit convection causing stars to inflate their radii (\citealt{Lopez-Morales2005}; \citealt{mullan01}). Another hypothesis suggests metallicity dependency on radii inflation: \cite{berger06} in their study find that the disagreement is larger among metal-rich stars than metal-poor stars. They conclude that current atmospheric models do not take into account opacity sources which may lead to such discrepancy. Therefore, it becomes imperative to discover more such systems and determine their masses and radii to very high precision.

F-type stars are typically fast rotators with a tenuous convective zone. \cite{Bouchy2011} suggest that there is a higher probability of M dwarfs orbiting F-type primaries, in contrast to G-type primaries. \cite{Bouchy2011,Bouchy2011b} further suggest that for a massive companion to exist around a primary star, the total angular momentum must be above a critical value. If a primary star has a smaller spin period than the orbital period of the system (as in the case for G-type stars), the tidal interactions between the two stars will cause the secondary companion to be eventually engulfed by the primary. However, this is less likely to occur among fast rotating F-type stars which have weaker magnetic braking and can avoid the spin-down caused due to the tides raised by the massive secondary. This helps the companions around F-type stars to survive rapid orbital decay due to loss of angular momentum.
F-type primaries with M-type secondaries (hereafter F+M binaries) are often discovered in transit surveys, which are primarily designed to search for transiting planets (e.g. \cite{bouchy05}; \cite{beatty07}). Over the last few years a handful of F+M binaries have been discovered and their properties determined (e.g. \citealt{Pont 2005a}; \citealt{2005b}; \citealt{2006}; \citealt{fernandez09}). Nonetheless, every additional system discovered and analysed contributes more to our understanding of fundamental stellar properties making the sample of F+M binaries an important subset of stellar studies. 

The NASA \textit{STEREO} mission consists of two satellites in the heliocentric orbit that study the Sun and its environment. An imager on one of the satellites is being employed to study the variability of bright stars and to look for transiting exoplanets. Observations taken by the \textit{STEREO} spacecraft were analysed by \cite{wraight12} to search for low-mass eclipsing companions to bright stars with effective temperatures between 4000~K~and~7000~K and visual magnitude 6~$<$~V~$<$~12. The \textit{STEREO} Heliospheric imagers (HI) have a field of view of 20$\degr$$~\times~$20$\degr$ and a 2048~$\times~$2048 element CCD with a spectral response that peaks between 6300~--~7300~\AA~\citep{wraight12}. HD~213597 was one of the 9 candidates selected \citep{wraight12} where the radius of the secondary was estimated to be below $0.4~R_{\sun}$. HD~213597A, with a visual magnitude of 7.8, is an F-type star having a rotational velocity of 40~km~s$^{-1}$ \citep{nordstrom04} and a radius of 2.039~R$_{\sun}$ \citep{masana06}. Table \ref{properties} lists the properties of this star obtained from the literature. The earlier studies on this star suggest that it may host a low mass companion.

Here, we describe high resolution spectroscopic and photometric observations and the methods of analysis used to derive the physical parameters concerning HD~213597 system. Details of the spectroscopic observations and transit photometry are reported in \S 2, while \S 3 describes the analytic methods and main results. Discussion and conclusions are presented in \S 4.

\section{Spectroscopic and Photometric Observations of HD~213597}

\subsection{Spectroscopy}
High-resolution spectroscopic observations of the star HD~213597A were made with the fiber-fed high-resolution spectrograph \citep[hereafter HRS;][]{tull98} at the 10-m McDonald Hobby-Eberly Telescope \citep[hereafter HET;][]{ramsey98}, and the fiber-fed echelle spectrograph, Physical Research Laboratory Advanced Radial velocity Abu-sky Search, \citep[hereafter PARAS;][]{chakraborty13} at the 1.2-m telescope at Gurushikhar Observatory, Mount Abu, India.

\subsubsection{HET-HRS observations}
We obtained 9 observations between 2011 October and 2012 July using the HRS in its 316g5936 cross-disperser setting, 2$\arcsec$ diameter fiber, resolution of $\sim$~30,000, and a simultaneous wavelength coverage of 4300~--~5800 \AA~and 6200~--~7600 \AA. The exposure times ranged between 60~s and 300~s, which yielded signal-to-noise ratio (SNR) between 200 and 600 per resolution element. Each observation was bracketed before and after with a Thorium-Argon (ThAr) lamp exposure for wavelength calibration. The same procedure was used on HD~215648, which served as a template to derive RVs. HD~215648 has a spectral type of F7V with $T_{\rm{eff}}$~=~6228~$\pm$~100~K, ${\rm log}~g$~=~4.15~$\pm$~0.07~\citep{edvardsson93}, $v \sin i$~=~6.7~$\pm$~0.7~km~s$^{-1}$ \citep{ammler12}, and $[Fe/H]$~=~$-$0.22 \citep{Valenti05}. The data were reduced using a custom optimal extraction pipeline as described in \cite{bender12}. 

\subsubsection{Mt Abu-PARAS observations}
A total of 15 observations were taken between October and November 2012 using the fiber-fed PARAS spectrograph which has 
a single prism as a cross disperser, a blaze angle of 75\degr, and a resolution of 67,000. The exposure time for each observation was fixed at 1200~s which resulted in SNR between 20 and 25 per pixel at the blaze wavelength. Simultaneous exposures of the science target and the ThAr lamp for wavelength calibration were taken. The wavelength region between 3700 and 6800 \AA~was considered for the RV measurements. The data were reduced by the Interactive Data Language (IDL) pipeline designed specifically for PARAS as described in \cite{chakraborty13}.

\subsubsection{Errors on Radial Velocity Measurements}
The precision on radial velocity (RV) measurements \citep{Hatzes92} is given as $\sigma~\sim~1.45~\times~10^9~(S/N)^{-1}~R^{-1}B^{-1/2}$~m~s$^{-1}$ where $S/N$ is the signal-to-noise of the spectra, while $R$ and $B$ are the resolving power and wavelength coverage of the spectrograph in angstrom~(\AA) respectively. Although PARAS has much higher resolution than the HET-HRS mode that was used, the SNR of PARAS spectra is lower, leading to the PARAS RVs having larger uncertainties than the HET-HRS RVs. As mentioned earlier, HD~213597A is an early F-type star with a rotational velocity of 40~km~s$^{-1}$. \cite{bouchy01} defined a quality factor Q that represents the quality and spectral line richness of the spectrum. It was further shown that Q deteriorates with increasing rotational velocity thereby increasing the RV uncertainty. For instance, a F2-type star with a rotational velocity of 40~km~s$^{-1}$ has a RV uncertainty 10 times larger than one that is rotating at 4~km~s$^{-1}$. The covariant errors obtained from fitting the peak of the cross correlation function in IDL are reported as uncertainties on the HET data. For PARAS spectra, errors based on photon noise are computed for each spectra as discussed in \cite{bouchy01}. Since HD~213597A has a RV semi-amplitude of $\sim$~33~km~s$^{-1}$, for a relatively small orbital period of 2.4238~d, the RVs get smeared within the duration of the exposure as a function of orbital phase. The orbital smearing errors were estimated and added quadratically with the photon noise and covariant errors for both PARAS and HET data points and used as uncertainties for individual RV points. The barycentric corrected RV values along with the associated uncertainties are mentioned in Table~\ref{rv}. Over the course of one year (October 2011 -- November 2012) we obtained a combined total of 24 observations with the two spectrographs. Two epochs from the observed radial velocities of PARAS data were removed in the fitting routine because of very low SNR (due to passing clouds). 

\subsection{Transit Photometry}
We performed ground-based photometric observations for this star from the calculated mid-eclipse time based on the ephemeris of \cite{wraight12}. Given the celestial coordinates of the star (Refer Table~\ref{properties}) and the prolonged monsoon in India, we get a narrow window of three weeks in the month of October to observe the complete transit of the object. Thus, we also revisited the archival \textit{STEREO} data \citep{wraight12}, in order to compute the transit parameters like transit duration and angle of inclination in combination with our ground-based photometry at PRL, Mount Abu, India.

\textit{STEREO} data from HI-1A instrument were extracted from the UK Solar System Data Centre (UKSSDC) website~\footnote{www.ukssdc.rl.ac.uk}. The star was observable for a period of 16 to 17 days on the CCD for each cycle of observation. Six such cycles of data were used for HD~213597A between the period January 2008 to October 2012. For our purposes, we used L2 images, which were pre-processed (bias-subtracted and flat-fielded) and accounted for solar coronal contamination. The wavelength band of observation is narrowly peaked between 630 nm and 730 nm \citep{wraight12}. A total of 36 transits for HD~213597A were recorded in 6 cycles of the extracted data.

We also obtained follow-up photometric observations of HD~213597A with the Physical Research Labortory (PRL) 10-inch telescope at Gurushikhar, Mount Abu, India. We carried out photometry on 2013 October 21 UT for a duration of 5~h. The observations were carried out using a Johnson \textit{R}-band with a back thinned E2V 1k~$\times$~1k CCD array with a field of view of 35\arcmin~$\times$~35\arcmin. A mismatch of the dome position while guiding resulted in the rejection of an hour-long observation. Furthermore, due to passing clouds, we lost about 20 per cent of the egress time. Despite the data loss, we detected the transit at a confidence level of $4\sigma$.

\section{Analysis and Results}

\subsection{Radial Velocity Measurements}

\subsubsection{HET-HRS}
The reduced data were wavelength calibrated and continuum normalized while the echelle orders were stitched to produce a continuous 1-D spectrum between two wavelength regions of 4300~--~5800 ~\AA~and 6200~--~7500 ~\AA. The spectrum was then divided into 8 segments (4386~--~4486~\AA; 4593~--~4843~\AA; 4925~--~5025~\AA; 5100~--~5410~\AA; 5475~--~5800~\AA; 6365~--~6430~\AA; 6620~--~6850~\AA; 7450~--~7500~\AA) that are free of telluric lines. Each segment was cross-correlated using a 1D cross-correlation algorithm. The resulting 1D correlation arrays were combined using the maximum likelihood method \citep{zucker03}. We employed a robust non-linear least square curve fitting algorithm, MPFIT in IDL, to fit the peak of the cross-correlation function and determine the RVs and the associated errors.

\subsubsection{Mt Abu-PARAS}
The entire reduction and analysis for PARAS was carried out by the custom-designed pipeline in IDL based on the REDUCE optimal extraction routines of \cite{Piskunov2002}. The pipeline performs the routine tasks of cosmic ray correction, dark subtraction, order tracing, and order extraction. A complete thorium line list for the PARAS spectral range is utilized. The wavelength calibration algorithms were specifically optimized for PARAS as described in \cite{chakraborty13}. The RV values were computed by cross-correlating the observed spectra against an F-type binary mask. Barycentric corrections were applied to all the RV points by standard IDL routines. The binary mask was created with the SPECTRUM program \citep{gray2009} using Kurucz stellar atmosphere models with a temperature of $T_{\rm eff}~=~6750$~K, ${\rm log}~g$~=~4.0 and a metallicity, $[M/H]$~=~$-0.2$ (refer Table~\ref{properties}).

\subsection{Spectral Analysis}
We based our spectral analysis of HD~213597A on the SME \citep{valenti96} spectral synthesis code. SME is composed of a radiative transfer engine that generates synthetic spectra from a given set of trial stellar parameters and a Levenberg-Marquardt solver that finds the set of parameters (and corresponding synthetic spectrum) that best matches observed input data in specific regions of the spectrum. The basic parameters that we used to define a synthetic spectrum are effective temperature ($T_{\rm eff}$), surface gravity (${\rm log}~g$), metallicity ($[M/H]$), iron abundance relative to solar ($[Fe/H]$) and projected rotational velocity ($v \sin i$). In order to match an observed spectrum, we solved for these five parameters. SME was implemented by using the Advanced Computing Centre for Research and Education (ACCRE) High-Performance Computing Center at Vanderbilt University for a large number (150) of different initial conditions to fully explore the $\chi^{2}$ space and find the optimal solution at the global minimum. In addition, we applied a line list based on \cite{stempels2007} that is suited for hotter stars, used the MARCS model atmosphere grid in the radiative transfer engine, and obtained the microturbulence ($v_t$) from the polynomial relation defined in \cite{gomez13}. 

We applied our SME pipeline to the high SNR HET spectrum of HD~213597A allowing all 5 parameters listed above to vary freely. We obtained $T_{\rm eff}$~=~$6625~\pm~121$~K, ${\rm log}~g$~=~$3.72~\pm~0.22$, and $[Fe/H]$~=~$-0.095~\pm~0.08$. It is important to note that for hotter stars, uncertainties on the gravity derived from spectral synthesis increase because the wings of the Mg~$\textsc{I}$~b triplet at 5167, 5172, and 5183~\AA~used to constrain this parameter become narrower and less sensitive to gravity. We also ran our SME pipeline again, this time constraining ${\rm log}~g~=~3.96~\pm~0.1$ based on the literature cited value. In this set of 150 trials, we fixed the gravity to a randomly chosen value within this range and solve for the other 4 parameters. For the constrained run, we obtained $T_{\rm eff}$~=~$6752~\pm~52$~K, iron abundance ($[Fe/H]$)~=~$-0.025~\pm~0.05$ and metallicity ($[M/H]$)~=~$-0.105~\pm0.03$.

\begin{figure}
\centering
\hbox{
\includegraphics[width=180mm]{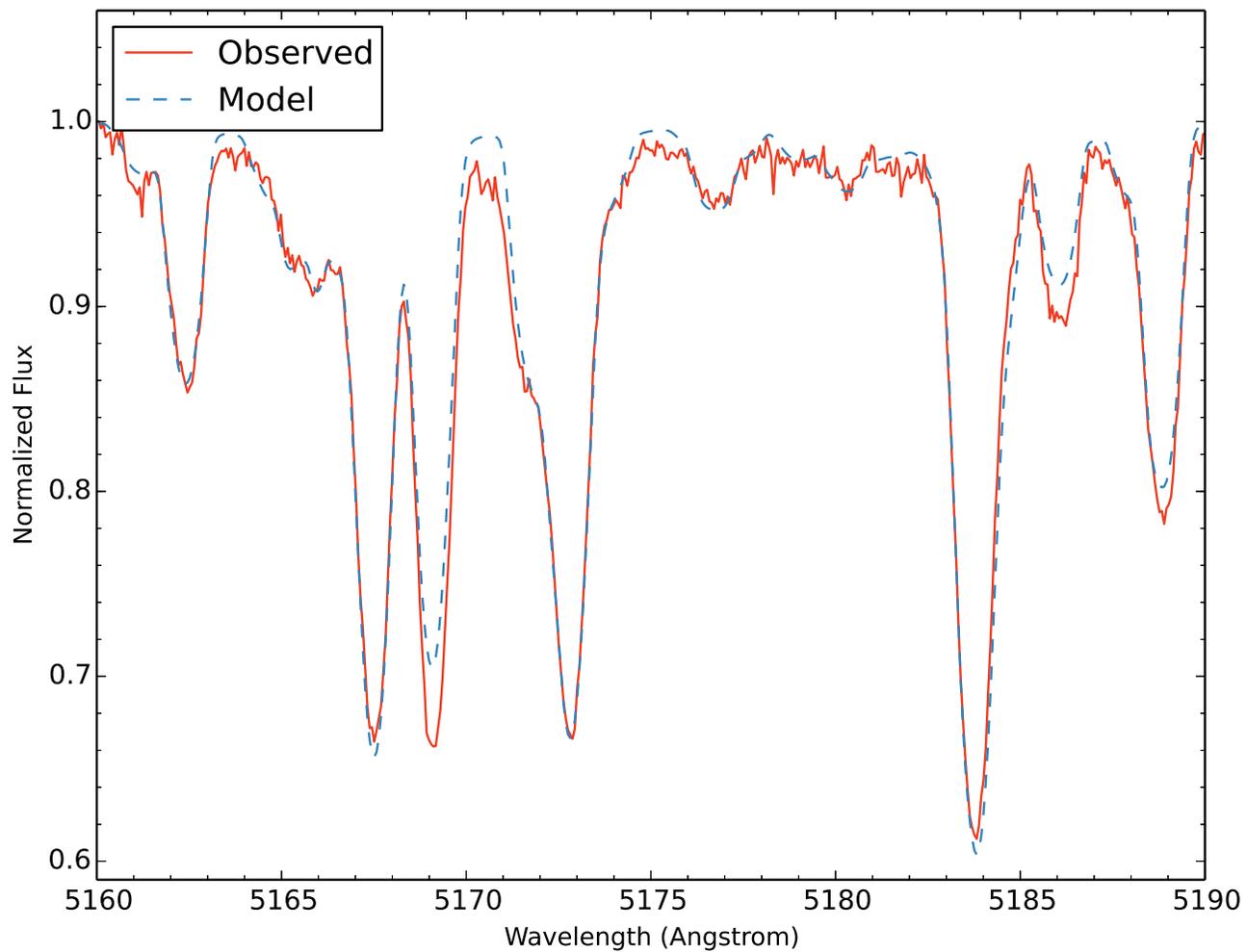}
}
\caption{Observed normalized spectra (solid line) plotted across the wavelength region of 5160--5190~\AA~covering the Mg I triplet at 5167, 5172, and 5183~\AA. Overplotted is the modelled spectra (dash line) obtained from SME analysis (when all parameters are kept free), with temperature value of $T\rm{_{eff}}$ of 6625 K, $[Fe/H]$ of $−0.095$ and ${\rm log}~g$ of 3.72.}
\label{sme}
\end{figure}

The formal 1-$\sigma$ errors are based on the $\delta \chi^{2}$ statistics for the 5 free parameters. To derive the systematic uncertainties, we compared the results of our pipeline to three independent datasets with parameters in the literature \citep{Valenti05,torres12,huber13} and report the mean absolute deviation of our results compared to all the comparison stars. The final results (both free and constrained ${\rm log}~g$) are listed in Table~\ref{smeparams} along with their combined statistical and systematic uncertainties. Fig.~\ref{sme} shows a sample of the observed spectrum (solid line) overlaid by the best-fit model obtained by a free parameter fit ($T_{\rm{eff}}$~=~6625~K; $[Fe/H]$~=~$-$0.095; ${\rm log}~g$~=~3.72; dotted line). The observed and model spectra are shown across the wavelength region 5160~--~5190~\AA, including the Mg~$\textsc{I}$~b triplet.

\subsection{Transit}
The \textit{STEREO} images from Heliospheric Imager, HI-1A, L2 data are a priori bias and flat corrected. These files have a field of view of 20$\degr~\times$~20$\degr$ imaged on a 2k~$\times$~2k CCD. The data were 2$~\times$~2 binned on a 1k~$\times$~1k image with a field of view of 70$\arcsec~\times$~70$\arcsec$ per pixel. Each image is a sum of 30 exposures with a total exposure time of 20~min and an observational cadence of 40~min \citep{wraight11}. The procedure given in \cite{sangaralingam11} was followed to do aperture photometry on the data. An aperture of 3.5~pixels was chosen for the same. We used the standard $\textsc{IRAF}$ \footnote{$\textsc{iraf}$ is distributed by the National Optical Astronomy Observatory, which is operated by the Association of Universities for Research in Astronomy, Inc., under cooperative agreement with the National Science Foundation.} $\tt{daophot}$ package for processing the photometry data. The flux computed by IRAF was detrended by fitting a 4$^{th}$ order polynomial to account for the CCD response function as discussed in \cite{sangaralingam11}. The data were further normalized for each cycle. The $\tt{rms}$ scatter on the light curve for the source star outside the transit time duration is 7~mmag. A total of 36 transits were obtained for the entire data of 6 cycles between the period January 2008 to October 2012.

For the ground-based photometry observations from Mt Abu on 21 October 2013, each individual frame had an exposure time of 2~s and a readout time of $\sim$~1~s. Similar to the \textit{STEREO} data, $\textsc{iraf}$ $\tt{daophot}$ package was used to perform aperture photometry on the data. The frames were flat-fielded and dark corrected. Varying air-mass, and other local sky variations are accounted for, to an extent, by performing differential photometry on a comparison star and a check star. For differential photometry, we need a comparison star which is non-varying with time and is similar in colour and magnitude to the program star. We chose HD~213763 with spectral type F5V and a visual magnitude of 7.8 as the comparison star. In order to cross check the non-variability of the comparison star, we required another field star called check star, which in our case was HD~213598 (K0V with a visual magnitude of 9.14). Although it is desirable to choose a check star of similar colour as that of the source and comparison stars, we settled for a K-type check star due to unavailability of bright field stars in the near vicinity. Since we used the Johnson \textit{R}-band for our photometry observations, the effect of scattering by moonlight and the atmospheric extinction was minimal. The average SNR of the source and the comparison star was between 300 and 350 for individual frames. Sixty consecutive frames, of similar exposures were median combined to avoid the scintillation noise from sky which, otherwise, becomes the dominant source of noise. Thus, each binned data frame comprised a total exposure time of ~120~s and a total time cadence of 202~s (including the read out time) making the SNR of the combined frame to be $\sim$~2500 for the same stars. Differential photometry was performed on the program star, HD~213597A and additional two bright field stars (HD~213763 and HD~213598). Individual light curve of comparison and check star showed a $\tt{rms}$ scatter of 12~mmag over the entire observation period. The $\tt{rms}$ scatter between the comparison and the check star reduced to 6~mmag after differential photometry. This formed the base level for detection of any transit and thus reflected the photometric error on each data point in the light curve.
The data from \textit{STEREO} and PRL 10-inch telescope were combined to produce a phase-folded light curve as both the datasets were observed in \textit{R}-band and had similar photometric errors on them.

\subsection{Simultaneous spectroscopy and photometry fitting}
We simultaneously fit the spectroscopy and photometry data with EXOFAST \citep{eastman2013}. EXOFAST is a set of IDL routines designed to fit transit and RV variations simultaneously or separately, and characterize the parameter uncertainties and covariances with a Differential Evolution Markov Chain Monte Carlo method \citep{johnson2011}. It can either fit RV and transit values exclusively or use both datasets to give a simultaneous fit. It also requires priors on $T_{\rm{eff}}$, ${\rm log}~g$ and iron abundance, $[Fe/H]$. EXOFAST uses empirical polynomial relations between masses and radii of stars; their ${\rm log}~g$, $T_{\rm{eff}}$, and $[Fe/H]$ based on a large sample of non-interacting binary stars in which all of these parameters were well-measured \citep{Torres2010}. These priors are used as a convenient way of modelling isochrones and are fast enough to incorporate them at each step in Markov chain. The errors derived here are determined by evaluating the posterior probability density based on the range of a given parameter that encompasses some set fraction of the probability density for the given model. We thereby caution the reader that the errors reported here for the mass and radius of the secondary are formal errors from EXOFAST. These values and errors are themselves based on models and isochrones and should not be used as independent observational checks on these models until more precise values can be derived in the future by teasing out a double lined spectroscopy orbital solution for this system.

\begin{figure}
\centering
\hbox{
\includegraphics[width=180mm]{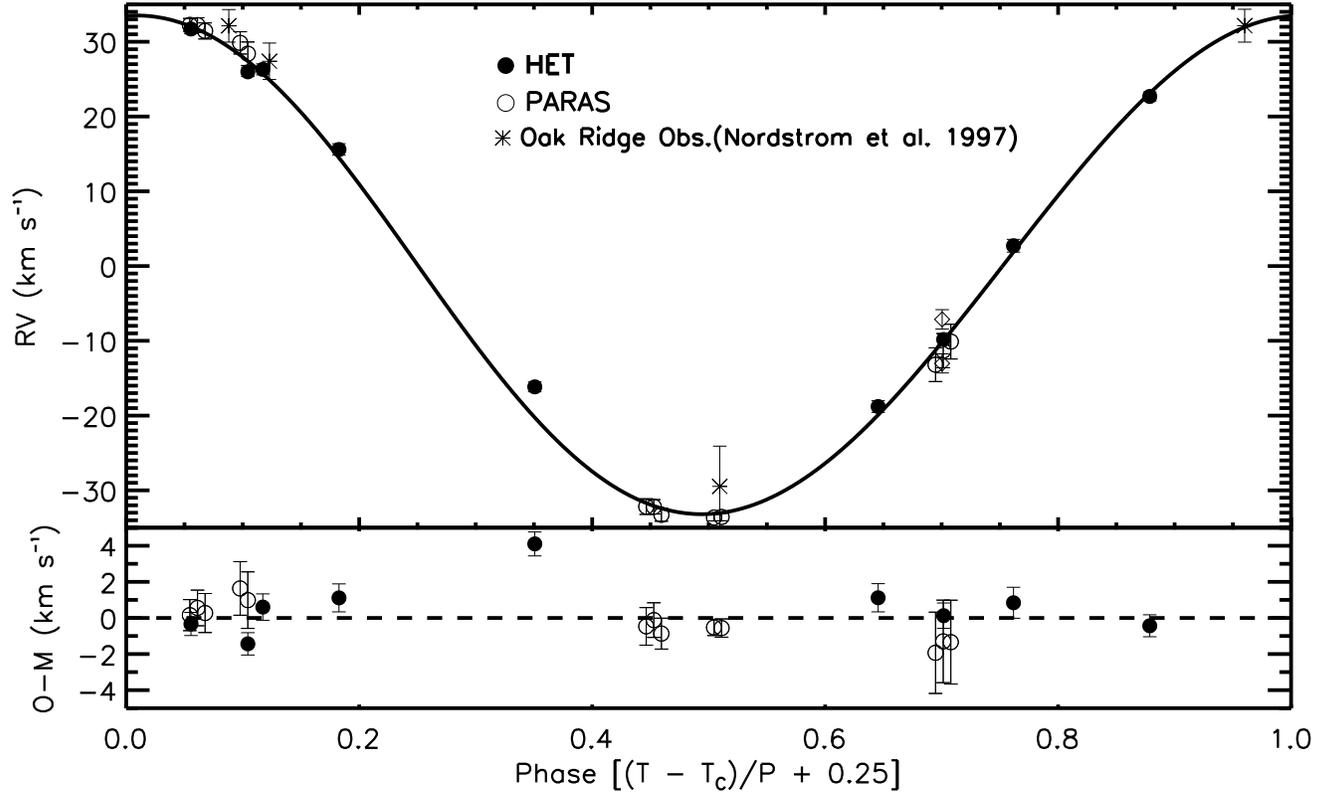}
}
\caption{(Top panel) RV model curve for star HD~213597 obtained from EXOFAST is plotted against orbital phase. HRS-HET (filled circles) and PARAS, Mount Abu (open circles) observed data points along with the estimated errors are overplotted on the curve. The two data points from PARAS not considered for RV fitting are overplotted with diamonds. The 4 RV measurements from Nordstr{\"o}m et al 1997 (not considered for RV fitting) are overplotted on the modelled data.
(Bottom panel) The residuals from best-fitting are plotted below the RV plot. The residuals are not plotted for the points which are not considered for the RV fitting.
For better visual representation, the x axis in Phase is shifted by 0.25 so that the central primary transit crossing point (Tc) occurs at phase 0.25 instead of 0.}
\label{RV_plot}
\end{figure}

We executed the EXOFAST routine with the combined RV datasets (HET-HRS and PARAS) and the combined photometry datasets (one from the transit light curve obtained by \textit{STEREO} and the other one obtained by PRL 10-inch telescope). On varying the SME-derived parameters as input priors to EXOFAST, we found that constraining ${\rm log}~g$ with the same uncertainties as given in Table~\ref{smeparams} yielded the most consistent values for $T_{\rm eff}$ and $[Fe/H]$ (best simultaneous fit for RV and transit data). The results of the execution are summarized in Table~\ref{result}. Fig.~\ref{RV_plot} illustrates the RV versus orbital phase for HD~213597A. Solid and open circles (top panel) show RV measurements of the primary taken with the HRS and PARAS instruments, respectively. The figure also shows the residuals (Observed-Model) in the bottom panel. There are four RV measurements for this star from \cite{nordstrom97} observed with the 1.5-m Wyeth reflector at the Oak Ridge Observatory in Harvard, Massachusetts. We corrected for the offset in RV and overplotted these points on Fig.~\ref{RV_plot} to give it a longer time base. However, we did not include these points in the model due to their relatively large errors. It is important to note that despite a long time-gap between our observations and those of \cite{nordstrom97}, their RV measurements (asterisks) lie on the model curve as shown in Fig.~\ref{RV_plot}. 

\begin{figure}
\centering
\hbox{
\includegraphics[width=180mm]{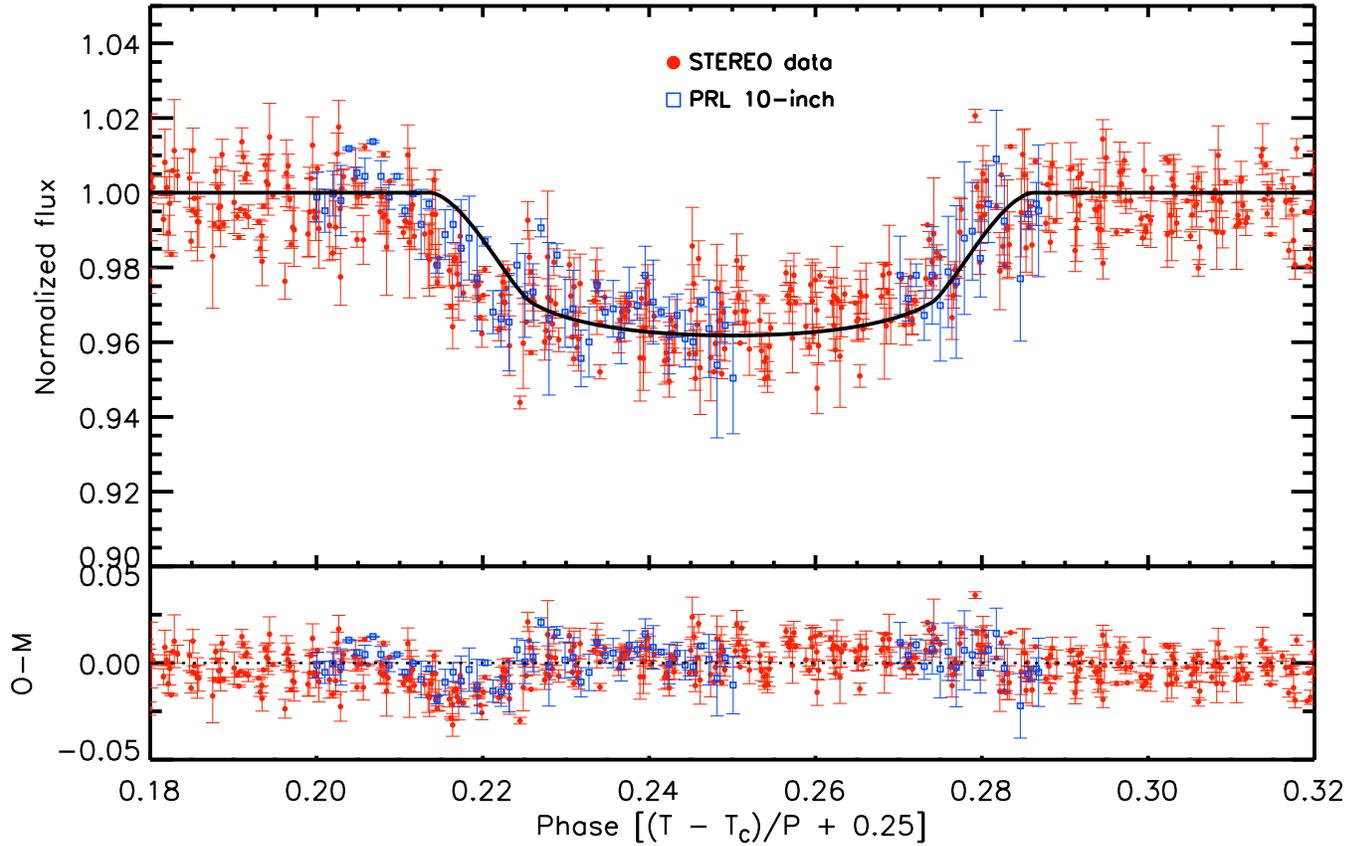}
}
\caption{(Top panel) Transit curve obtained from \textit{STEREO} data and PRL 10-inch telescope is plotted based on the parameters from EXOFAST. The \textit{STEREO} data are plotted with red filled circles and the PRL 10-inch telescope data are plotted with open blue squares along with their individual error bars. (Colours are only in online version.)
(Bottom panel) Observed -Fit residuals are plotted.
For better visual representation, the x axis in Phase is shifted by 0.25 so that the central primary transit crossing point (Tc) occurs at phase 0.25 instead of 0.}
\label{transit}
\end{figure}

Fig.~\ref{transit} (upper panel) shows the simultaneous fit for the transit light curve obtained by analysing \textit{STEREO} archival data (filled circles) and PRL 10-inch telescope data (open squares) overplotted with the model derived from EXOFAST (solid curve). The residuals are plotted in lower panel. The simultaneous fit gives us a transit depth of $0.0317_{-0.0012}^{+0.0013}$~mag, angle of inclination of $84.9\degr~_{-0.5}^{+0.62}$ and a transit time duration of $274_{-4}^{+5}$~min. The transit depth cited here is consistent with the refined analysis of \textit{STEREO} light curve by \cite{whittaker13}. We obtain the mass of the secondary as $0.286_{-0.01}^{+0.012}~M_{\sun}$ and a radius of $0.344_{-0.01}^{+0.097}~R_{\sun}$. We also calculated the mass function for this system by using the IDL BOOTTRAN package \citep{wang10}.
The results obtained by the combined RV and transit datasets are consistent with the period 2.4238~d \citep{wraight12,whittaker13} with a semi-major axis of 4.48~R$_{\rm star}$ (0.041~AU) and a RV semi-amplitude of 33.39~km~s$^{-1}$. Based on the mass limits, we conclude that the secondary star is an early M dwarf \citep{Baraffe1996}.

\section{DISCUSSION AND CONCLUSIONS}
Using high resolution spectroscopy taken with the HET-HRS, PARAS spectrographs and photometry with the \textit{STEREO} data, PRL 10-inch telescope, we conclude that HD~213597 is an eclipsing binary system with an F-type primary and an early M-type secondary. The estimated secondary mass from RV measurements is $M_B~=~0.286_{-0.01}^{+0.012}~M_{\sun}$ with an accuracy of $\sim~3-~4$~per~cent (formal errors). Based on the transit depth of $0.0317_{-0.0012}^{+0.0013}$~mag, the radius of the secondary is estimated to be $R_B~=~0.344_{-0.01}^{+0.097}~R_{\sun}$ with an accuracy of $\sim$~$5-6$~per~cent (formal errors). The radius estimated here is in good agreement with the value given by \cite{wraight12}. 

Metallicity and the iron abundance of the system are accurately determined in this work as $-0.105$~$\pm$~0.03 and $-0.025$~$\pm$~0.05 (see Table~\ref{smeparams}) by spectroscopic analysis. From the models for M dwarfs \citep{Baraffe1998}, we estimate the T$_{\rm{eff}}$ of the secondary star for a mass of $\sim~0.3~M_{\sun}$ (from this work) and an age of $\sim$~2~Gyr \citep{casagrande11}, as 3437~K for a $[M/H]$ of $0.0$, and 3643~K for a $[M/H]$ of $-0.5$. From \cite{Baraffe1998} models, the radius for $0.3~M_{\sun}$ turns out to be $0.29~R_{\sun}$ for \rm{$[M/H]~=~0.0$} and $0.28~R_{\sun}$ for \rm{$[M/H]~=~-0.5$}. Clearly, these theoretically derived radius values are lower than the observations presented here.

The binary system has an inclination angle of $84\degr.9_{-0.5}^{+0.61}$. The rotational velocity of the primary star obtained by SME analysis, as mentioned in Table~\ref{smeparams} is 40~km~s$^{-1}$, which matches when measured by the width of the cross-correlation function. This value is comparatively lower than the reported rotational velocities for similar F-type stars \citep{glebocki2005}. With the knowledge of rotational velocity of the primary star and its radius of $\sim~2.03~R_{\sun}$ from Table~\ref{properties}, the projected spin period, P/$\sin~i$ of the primary star is calculated to be 2.422 d, which is close to the observed orbital period of the system. The eccentricity of the system, calculated as 0.0198, makes it close to circular. Since the orbital and rotational periods for the star are synchronous, this star forms another evidence for tidal circularization playing role in a close binary system \citep{mazeh2008}.

The detailed spectral analysis of the high SNR HET spectra gives us $T_{\rm eff}$=~$6752~\pm~52$~K, $[Fe/H]$~=~$-0.025~\pm~0.05$ for  ${\rm log}~g$~=~$3.96~\pm~0.1$. The temperature indicates an F-type primary star. It may be mentioned that \cite{masana06} obtained a $T\rm{_{eff}}$ of 6936~$\pm$~70 K based on V and 2MASS \textit{JHK} photometry, while \cite{casagrande11} reported a value of 6837~$\pm$~80 K based on improved colour calibration of the same bands. Our results are more reliable as they are based on detailed spectral analysis from high SNR spectra. Furthermore, these results are substantiated by the fact that the RV and photometry data are not able to constrain a solution in EXOFAST by using the previously cited $T\rm{_{eff}}$ value of 6936~$\pm$~70 K. Using $T\rm{_{eff}}$, $[Fe/H]$ and ${\rm log}~g$ derived from this work, the mass and radius of the primary star obtained from Mass-Radius relationship \citep{Torres2010} are $1.48~M_{\sun}$ and $2.11~R_{\sun}$, respectively, consistent with the values reported in literature (see Table~\ref{properties}). 

Future spectroscopic observations of this star during transit will enable us to observe the Rossiter-McLaughlin effect \citep{scott2007}, and help determine whether the secondary star is in retrograde or prograde orbital motion with respect to the rotation of the primary. This may lead to a better understanding of the binary formation mechanisms at a primordial stage. Future high resolution near-infrared observations will be able to provide dynamical masses of the system.  

\section*{Acknowledgments}
This work has been made possible by the PRL research grant for PC (author) and the PRL -DOS (Department of Space, Government of India) grant for the PARAS spectrograph. This work was partially supported by funding from the Center for Exoplanets and Habitable Worlds. The Center for Exoplanets and Habitable Worlds is supported by the Pennsylvania State University, the Eberly College of Science, and the Pennsylvania Space Grant Consortium. We thank the observatory staff at the Hobby-Eberly Telescope (HET) and the PRL Mt Abu telescope from where the data were obtained. HET is a joint project of the University of Texas at Austin, the Pennsylvania State University, Stanford University, Ludwig-Maximilians-Universit{\"a}t M{\"u}nchen, and Georg-August-Universit{\"a}t G{\"o}ttingen. The HET is named in honor of its principal benefactors, William P. Hobby and Robert E. Eberly. PC acknowledges help from Varun Dongre and Vishal Shah for their technical support during the course of data acquisition. This research has made use of the ADS and CDS databases, operated at the CDS, Strasbourg, France. We also thank the referee, Barry Smalley, for detailed and insightful comments which have helped to improve the quality of the paper significantly.

%%%%%%%%%%%%%%%%%%%%%%%%%%%%%%%%%%%%

%------------------------------------------------------------------------------------------------
%  TABLE 1 
%------------------------------------------------------------------------------------------------
\newpage

\begin{table}

\caption{Stellar properties of host star HD~213597A from literature.}
\label{properties}
\begin{tabular}{llr}\\
%\begin{large}
\hline
\multicolumn{1}{c}{Parameter} & \multicolumn{1}{c}{Value} & \multicolumn{1}{c}{Reference} \\
\hline
Mass           &       1.5 $\pm$0.1 \rm${M\sun}$	   & 	(1)  \\
Radius         &       2.039$\pm$0.303 \rm${R\sun}$   &	(2)  \\
${\rm log}~g$         &       3.99$\pm$0.05		               &	(1)  \\
$v \sin i$     &       40$\pm$5 km s$^{-1}$                &	(3)  \\
$T\rm{_{eff}}$ &       6837$\pm$80    K     &    (1)    \\
$[Fe/H]$          &	      $-$0.14$\pm$0.1            &   (1) \\
Age	           &	      1.90$\pm$0.2 Gyr                     &(1) \\
Distance	   &	      115 $\pm$15 pc		             & (4) \\
V magnitude	&		7.8			&  (5) \\
RA(epoch=2000)	   &	$22^{h}$  $32^{m}$ $32^{s}.626$ & (5) \\
Dec(epoch=2000)		&	1\degr 34\arcmin 56\arcsec.83 & (5) \\	
\hline
%\end{large}
\end{tabular}\\
References: (1) \cite{casagrande11}; (2) \cite{masana06}; \\
(3) \cite{nordstrom04}; (4) from HIPPARCOS data  (5)from SIMBAD\\

\end{table}

\clearpage

%------------------------------------------------------------------------------------------------
%  TABLE 2
%------------------------------------------------------------------------------------------------

\newpage
\begin{table}
\small
\caption{Observations for the star HD~213597.
The time in UT and BJD-TDB are mentioned in first two columns followed by exposure time and barycentric corrected RV in third and fourth column respectively. The next two columns have RV errors, fifth column corresponds to covariant errors (HET-HRS) or photon noise errors (PARAS) and column 6 has orbital smearing errors. Last column is the instrument used for observations.}
\label{rv}
\begin{tabular}{lcccccc}\\
\hline
\multicolumn{1}{c}{UT Date} & \multicolumn{1}{c}{T-2,400,000} & \multicolumn{1}{c}{Exp. Time} & \multicolumn{1}{c}{RV} &\multicolumn{1}{c}{$\sigma$-RV} &\multicolumn{1}{c}
{$\sigma$-Orbital Smearing}&\multicolumn{1}{c}{Instrument}\\
     &	(BJD-TDB) & (sec.) &  (km s$^{-1}$)  &	(km s$^{-1}$) & (km s$^{-1}$)  &	(flag)	\\
\hline
2011 Oct  30 & 55864.689  & 250		& 4.844      & 0.209  & 0.167		 & HET-HRS \\
2011 Nov  25 & 55890.614  & 60		& 11.822     & 0.208  & 0.030		 &  HET-HRS \\
2011 Nov  27 & 55892.609  & 180 	& $-$20.967  & 0.214  & 0.116		& HET-HRS \\
2011 Dec  15 & 55910.552  & 60 		& 15.026     & 0.212  & 0.026 		& HET-HRS \\
2012 Jun  11 & 56089.949  & 300		& 14.561     & 0.210  & 0.142		 & HET-HRS \\
2012 Jun  22 & 56100.924  & 300		& $-$30.586  & 0.217  & 0.162		 & HET-HRS \\
2012 Jun  23 & 56101.918  & 300		& 19.909     & 0.212  & 0.064		 & HET-HRS \\
2012 Jul  01 & 56109.905  & 180 	& $-$27.993  & 0.204 	& 0.103		& HET-HRS \\
2012 Jul  02 & 56110.901  & 300		& $-$9.125   & 0.214 	& 0.203		 & HET-HRS \\
2012 Oct  17 & 56218.260  & 1200 	& 2.940      & 0.148  & 0.255	 &	 PARAS \\
2012 Oct  17 & 56218.276  & 1200	& 2.905      & 0.176  & 0.290  		& PARAS \\
2012 Oct  17 & 56218.292  & 1200	& 2.134      & 0.180  & 0.325		 & PARAS \\
2012 Oct  18 & 56219.210  & 1200	& $-$61.491  & 0.248  & 0.253		 & PARAS \\
2012 Oct  18 & 56219.225  & 1200 	& $-$61.522  & 0.242  & 0.219		 & PARAS \\
2012 Oct  18 & 56219.241  & 1200	& $-$62.592  & 0.226  & 0.185		 & PARAS \\
2012 Oct  21{$^{\star}$}  & 56222.269 & 1200	 & $-$42.356  & 0.267  & 0.805  & PARAS\\
2012 Oct  21{$^{\star}$}  & 56222.285 & 1200	  & $-$36.462  & 0.283  & 0.813  & PARAS \\
2012 Oct  23 & 56224.198  & 1200	& $-$62.984  & 0.148  & 0.051 		 & PARAS \\
2012 Oct  23 & 56224.214  & 1200	& $-$62.904  & 0.158  & 0.085		 & PARAS \\
2012 Nov   7 & 56239.202  & 1200	& $-$42.605  & 0.148  & 0.763		 & PARAS \\
2012 Nov   7 & 56239.219  & 1200 	& $-$40.721  & 0.155  & 0.773		& PARAS \\
2012 Nov   7 & 56239.234  & 1200	& $-$39.516  & 0.167  & 0.783		 & PARAS \\
2012 Nov   8 & 56240.179  & 1200	& 0.438     & 0.188  &  0.479		 & PARAS \\
2012 Nov   8 & 56240.195 & 1200 	 & $-$1.008   & 0.183  & 0.511		& PARAS \\
\hline
\end{tabular}\\
{$^{\star}$}Not considered for RV analysis
\end{table}
\clearpage

%------------------------------------------------------------------------------------------------
%  TABLE 3
%------------------------------------------------------------------------------------------------

\begin{table}
\begin{center}
\caption{Spectroscopically determined stellar parameters of HD ~213597A (this work). The two measurements listed for each stellar parameter results from two SME run: (1) when all parameters are allowed to float; (2) when ${\rm log}~g$ is fixed to a value determined by mass and radius measurements.}
\label{smeparams}
\begin{tabular}{lcccc} \\
\hline
\multicolumn{1}{c}{Stellar Parameter} & \multicolumn{1}{c}{Value (Free parameters)} & \multicolumn{1}{c}{Uncertainty} & \multicolumn{1}{c}{Value (Constrained fitting)} & \multicolumn{1}{c}{Uncertainty}  \\

$T\rm{_{eff}}$~(K)      &   6625  & $\pm$  121 & 6752  & $\pm$  52  \\
${\rm log}~g$          &   3.72  & $\pm$  0.22 &3.96  & $--$ \\
$[Fe/H]$           &   -0.095 & $\pm$ 0.08 & -0.025 & $\pm$ 0.05\\
$[M/H]$		&      -0.156 & $\pm$ 0.03 & -0.105 & $\pm$ 0.03\\
$v \sin i$ (~km~s$^{-1}$)  &   39.53 & $\pm$ 1.3 & 39.53 & $\pm$ 1.3\\ 
Microturbulence (~km~s$^{-1}$) &  2 & $--$ & 2   & $--$  \\
\hline 
\end{tabular}\\
\end{center}
\end{table}

%------------------------------------------------------------------------------------------------
%  TABLE 4
%------------------------------------------------------------------------------------------------
\clearpage
\newpage
\begin{table}
\small
\caption{Results obtained from EXOFAST by simulataneous fitting of RV and transit data for HD~213597 with a 68\% confidence interval.} 
\label{result}
\begin{tabular}{lcc}\\
\hline
\multicolumn{1}{c}{~~~Parameter} & \multicolumn{1}{c}{Units} & \multicolumn{1}{c}{Values}  \\ 
\hline
{HD~213597A:} &  \\

                           ~~~$M_{A}$\dotfill &Mass (M$\sun$)\dotfill & $1.293_{-0.1}^{+0.12}$  \\
                         ~~~$R_{A}$\dotfill &Radius (R$\sun$)\dotfill & $1.973_{-0.080}^{+0.085}$ \\
                     ~~~$L_{A}$\dotfill &Luminosity (L$\sun$)\dotfill & $7.04_{-0.52}^{+0.60}$ \\
                         ~~~$\rho_{A}$\dotfill &Density (cgs)\dotfill &$0.2377_{-0.0053}^{+0.0054}$ \\
              ~~~${\rm log}~g_{A}$\dotfill &Surface gravity (cgs)\dotfill &$3.96\pm0.01$ \\
              ~~~T$_{\rm{eff}}$  &Effective temperature (K)\dotfill &$6699_{-52}^{+53}$\\
                              ~~~$[$Fe$/$H$]$ & Iron Abundance\dotfill &$-0.065_{-0.048}^{+0.050}$ \\
{HD~213597B:} &  \\
                               ~~~$e$\dotfill &Eccentricity\dotfill &$0.0198_{-0.0091}^{+0.0094}$\\
    ~~~$\omega_*$\dotfill &Argument of periastron (degrees)\dotfill &$72_{-18}^{+12}$\\
                              ~~~$P$\dotfill &Period (days)\dotfill &$2.4238503_{-0.0000019}^{+0.000017}$\\

~~~$a$\dotfill &Semi-major axis (AU)\dotfill & $0.04112_{-0.00057}^{+0.00058}$ \\
                              ~~~$M_{B}$\dotfill &Mass (M$\sun$)\dotfill & $0.286_{-0.01}^{+0.012}$ \\
                           ~~~$R_{B}$\dotfill &Radius (R$\sun$)\dotfill & $0.344_{-0.01}^{+0.097}$  \\
                       ~~~$\rho_{B}$\dotfill &Density (cgs)\dotfill & $9.32_{-0.41}^{+0.43}$ \\
                  ~~~${\rm log}~g_{B}$\dotfill &Surface gravity\dotfill & $4.803_{-0.0097}^{+0.0093}$  \\

{RV Parameters:} &  \\          

                              ~~~$e\cos\omega_*$\dotfill & \dotfill & $0.0056_{-0.0042}^{+0.0043}$ \\
                              ~~~$e\sin\omega_*$\dotfill & \dotfill & $0.0184_{-0.0097}^{+0.0096}$ \\
           ~~~$T_{P}$\dotfill &Time of periastron (BJD)\dotfill & $2455392.09_{-0.12}^{+0.087}$ \\
                    ~~~$K$\dotfill &RV semi-amplitude m~s$^{-1}$\dotfill & $33390_{-280}^{+290}$ \\
                   ~~~$M_{B}/M_{A}$\dotfill &Mass ratio\dotfill & $0.2218\pm0.0045$ \\
                   ~~~$\gamma$ \dotfill& Systemic RV m~s$^{-1}$ \dotfill & $-4420_{-190}^{+180}$ \\
                  ~~~$f(m)$ \dotfill & Mass function {$^{\star}$} (M$\sun$) \dotfill & $0.0091 \pm 0.000048$  \\

{Transit Parameters:} &  \\             

                ~~~$T_C$\dotfill &Time of transit (BJD)\dotfill & $2455392.2014_{-0.0016}^{+0.0015}$ \\
~~~$R_{B}/R_{A}$\dotfill &Radius of secondary in terms of primary radius\dotfill & $0.178\pm0.001$ \\
 ~~~$i$\dotfill &Inclination (degrees)\dotfill & $84.9_{-0.50}^{+0.61}$ \\
     ~~~$a/R_{A}$\dotfill &Semi-major axis in stellar radii\dotfill & $4.48_{-0.038}^{+0.039}$ \\
       ~~~$\delta$\dotfill &Transit depth\dotfill & $0.0317_{-0.0012}^{+0.0013}$ \\
       ~~~$T_{14}$\dotfill &Total transit duration (minutes)\dotfill & $274_{-4}^{+5}$ \\
\hline
\end{tabular}\\
{$^{\star}$} \small {By BOOTTRAN (IDL)}. 
\end{table} 
   
%------------------------------------------------------------------------------------------------
%  FIG 1
%------------------------------------------------------------------------------------------------

\newpage
\clearpage

%------------------------------------------------------------------------------------------------
%  FIG 2
%------------------------------------------------------------------------------------------------

\newpage
\clearpage
%------------------------------------------------------------------------------------------------
%  FIG 3
%------------------------------------------------------------------------------------------------

\newpage
\clearpage

\end{document}